\input{aipcheck}

\documentclass[
    ,final            
  ]
  {aipproc}

\layoutstyle{6x9}

\usepackage{xspace}

\begin{document}

\title{Searches for SUSY with the ATLAS detector}


\author{Osamu Jinnouchi, on behalf of the ATLAS collaboration}{
  address={Tokyo Institute of Technology, O-okayama, Meguro-ku, Tokyo
  152-8851 Japan}
}
\def\met{\ensuremath{E_{\mathrm{T}}^{\mathrm{miss}}} \xspace} 
\def\pbinv{\ensuremath{\mathrm{pb}^{\mathrm{-1}}} \xspace} 
\def\fbinv{\ensuremath{\mathrm{fb}^{\mathrm{-1}}} \xspace} 

\def\pT{\ensuremath{p_{\mathrm{T}}} \xspace} 
\def\ET{\ensuremath{E_{\mathrm{T}}} \xspace} 
\def\MT{\ensuremath{M_{\mathrm{T}}} \xspace} 
\def\TeV{\ifmmode {\mathrm{\ Te\kern -0.1em V}}\else
                   \textrm{Te\kern -0.1em V}\fi \xspace}%
\def\GeV{\ifmmode {\mathrm{\ Ge\kern -0.1em V}}\else
                   \textrm{Ge\kern -0.1em V}\fi \xspace}%
\def\MeV{\ifmmode {\mathrm{\ Me\kern -0.1em V}}\else
                   \textrm{Me\kern -0.1em V}\fi \xspace}%

\begin{abstract}
We present a review of the SUSY search strategies in ATLAS in conjunction 
with a readiness of the detector systems for first collision data in
2009 fall. Commissioning was performed with the LHC single beams and the
 cosmic ray data in 2008. The talk covers the
analysis strategies based on the large \met plus high \pT multi-jets 
signature in which the number of methods are investigated to extract background
estimation from real data. The expected discovery reach with inclusive 
analysis is shown. The review also covers the special signature searches 
for certain SUSY scenarios, where specific detector components
play a crucial role in detecting and measuring them. 
\end{abstract}

\maketitle


\section{Introduction}
The world's most exciting moment is finally about to happen this year
2009. The Large Hadron Collider (LHC) at CERN is expected to deliver a
few 100 \pbinv of integrated luminosity to experiments in the first year(s).
The ATLAS collaboration is now ready for the first $pp$ collision data 
from LHC, after 15 years of preparation and integration of
the detector systems. The theoretical potential for discovering  
physics beyond the Standard Model (SM) is quite high, especially the
supersymmetry (SUSY) is one of the highly motivated scenarios 
expected to be discovered at LHC. 
It is vital to corroborate our final readiness before the 
dawn of new physics, on both the detector systems and the analysis 
strategies.

At LHC energy regime, the SUSY production process is dominated by
the strongly interacting particles, namely the squarks and gluinos,
which are typically the heaviest among the SUSY particles. The cross section 
is fairly independent from the detail of phenomenology models behind. This 
allows us to generalize our SUSY search strategies. 
In the context of Minimal Supersymmetric extension of the Standard 
Model (MSSM) with R-parity conservation, these pair-produced 
SUSY particles (majority of which are colored)
decay to lighter particles. They cascade down until 
they reach the lightest SUSY particle (LSP). 
Therefore rather complex event signatures are 
typically expected. The details of the cascade decays are functions 
of model parameters, which cannot currently be predicted. 
Our strategies should rely on the robust signatures which cover 
large classes of models, and in the same time it should be clearly 
distinctive from the SM backgrounds. Therefore, in our base analysis 
we look for the signatures as,
\begin{equation}
\mbox{large }\met+\mbox{high \pT multi-jets}+(\mbox{leptons}, b\mbox{-jets}, \tau\mbox{-jets})
\end{equation}
and in some cases, photons or long-lived SUSY particle signatures 
exist on top of these basic requirements. Obviously good 
performance of the detector ingredients is the key for the early 
confirmation of SUSY. 
The questions we must ask are: (i) Are the detectors 
ready for the first collision data? Do they achieve expected performance 
in commissioning? (ii) Are we able to get 
SM backgrounds under control? Could we estimate the background without 
relying too much on the Monte Carlo? (iii) 
If special signatures exist on top of the basic requirements, 
are we able to detect them?
The current answers for these questions are addressed in this talk.
\section{Status of ATLAS detectors in commissioning}
The ATLAS detector is a multi-purpose detector systems composed of
4 major components: inner trackers, calorimeters, muon
spectrometers and magnets, whose detailed description can be found 
elsewhere \cite{ATLAS-detector} (and references therein). 
The most important tasks which have to be carried out in the
commissioning phase are the calibrations and the alignments. These 
have to come before any physics measurements. 
There are numbers of in-situ calibration/alignment menus scheduled 
with the early physics data which one can address when the collision
data arrives. For instance, we use  
$Z\rightarrow ee$, $J/\Psi$ samples to determine the $e/\gamma$ 
energy scale . 
However even before arrival of collision data,
the current commissioning programs on
cosmic ray data, single beam data from 2008, and the stand-alone 
calibration/alignment systems, can address the performance on 
track alignments, azimuthal asymmetry/uniformity of Calorimeters,
muon system alignment, etc. 
\paragraph{Commissioning with cosmic rays}
The activity started as early as 2005 in parallel with the detector 
installation. In the last few years, the work evolved from single 
component operations to combined detector running. With significant 
number of events, we were able to gain experience in detector 
operation and control, DAQ and analysis chain, together with the 
understanding of in-situ performance. Integrated cosmic data from 
the combined commissioning since Sept. 2008, exceeded 200M events,
allowing fairly precise performance control.       
\paragraph{Commissioning with single beam}
\begin{figure}
\includegraphics[height=.23\textheight]{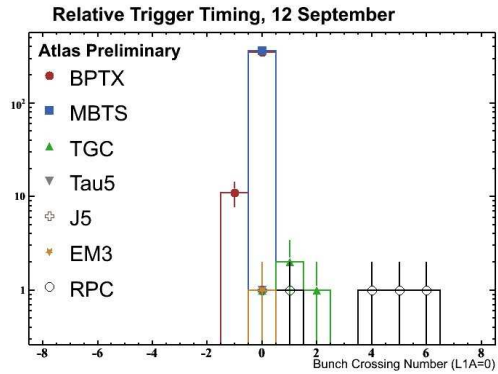}
\caption{Timings of various trigger menus at the 3rd day of the Sept.'08 
commissioning, relative to the MBTS timing in unit of bunch crossings.}
\label{fig:TriggerTiming}
\end{figure}
As described in more detail in \cite{PJenni}, the first proton 
beams in LHC rings were circulated for ten days starting on Sept. 
10th 2008. ATLAS made a good use of this opportunity for timing 
adjustments for trigger and detector system. 
Trigger timing was firstly adjusted using Minimum Bias 
Trigger Scintillators (MBTS) located at the surface of End-cap 
electromagnetic Calorimeters, and the beam timing pick-up (BPTX) 
located at upstream beam line 175$m$ away from ATLAS. 
Figure~\ref{fig:TriggerTiming} shows the situation on the third day, 
where various LVL-1 trigger timings are shown with respect to the MBTS. 
Relative timing between MBTS and BPTX, and the beam quality had been 
significantly improved in a short time. Also with
so-called ``splash'' events (beam dump collimator were placed 140$m$ 
upstream from ATLAS),
one could align the timing of large detector volumes within 1 $ns$.  
\paragraph{Readiness of the inner detector}
\begin{figure}
\includegraphics[height=.23\textheight]{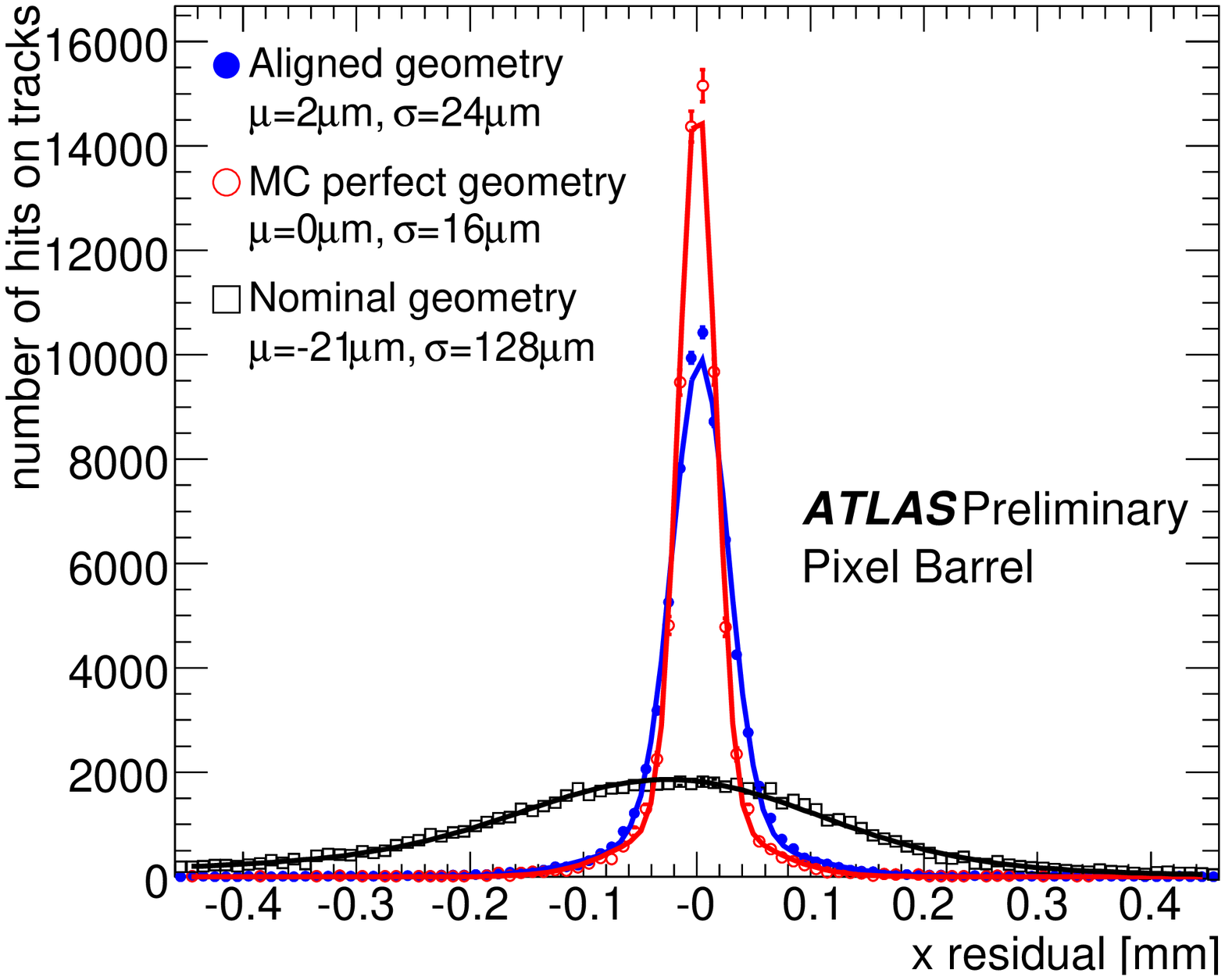}
\includegraphics[height=.23\textheight]{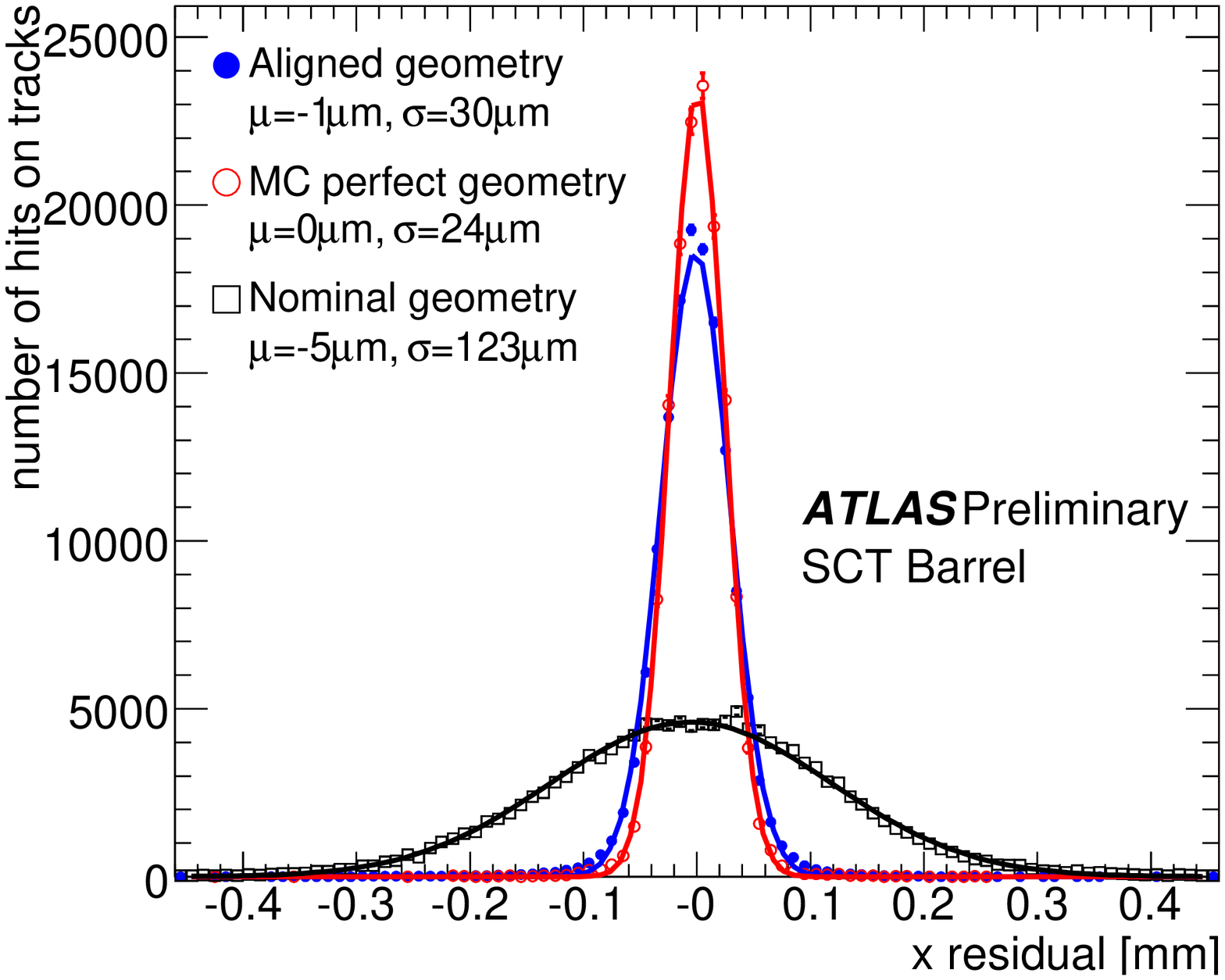}
\caption{Residual distributions of inner detectors for cosmic ray data; 
(left) Pixel tracker for precision coordinate in barrel 
region, (right) SCT for precision coordinate in barrel region. 
(red) distribution from MC
simulation obtained with perfect geometry: (black) before the alignment, 
(blue) after the alignment
 }
\label{fig:IDresidual}
\end{figure}
The alignment of the ATLAS Inner detector components (from inside: Pixel, SCT, TRT)
is a crucial task in reaching the design performance. The procedure is
based on the minimization of hit residuals for high \pT tracks, 
e.g. cosmic muons. Figure~\ref{fig:IDresidual} shows the residuals 
for Pixel ($\sigma=24\mu m$) and SCT ($\sigma=30\mu m$). The residual 
widths after the alignment are quite close to the performance which 
one could expect with perfect geometry in simulation. The same is true
for the TRT. The fraction of masked channels due to the hardware failure 
is low, Pixel (well below 0.02\%), SCT ($\sim$1\% in Barrel, 
$\sim$3\% in End-cap, aiming $<1\%$ in 2009). The efficiency is high, 
i.e. Pixel($\sim99.8\%$), SCT($>\sim99.0\%$).
With the TRT, the transition radiation from high \pT muon ($>100$\GeV) is also
observed, the rate agrees well with the test beam result.    
\paragraph{Readiness of the Calorimeters}
\begin{figure}
\includegraphics[height=.23\textheight]{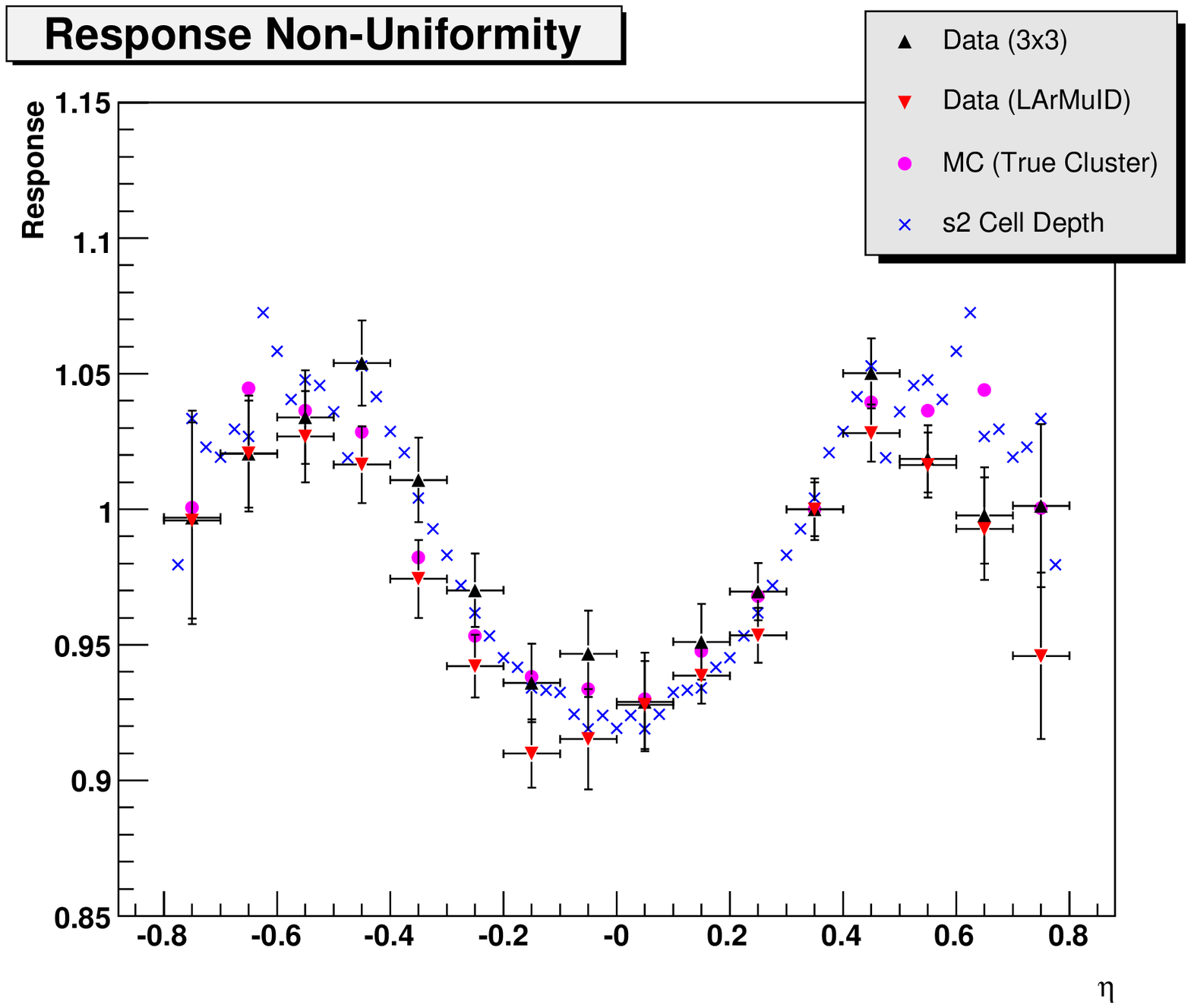}
\includegraphics[height=.23\textheight]{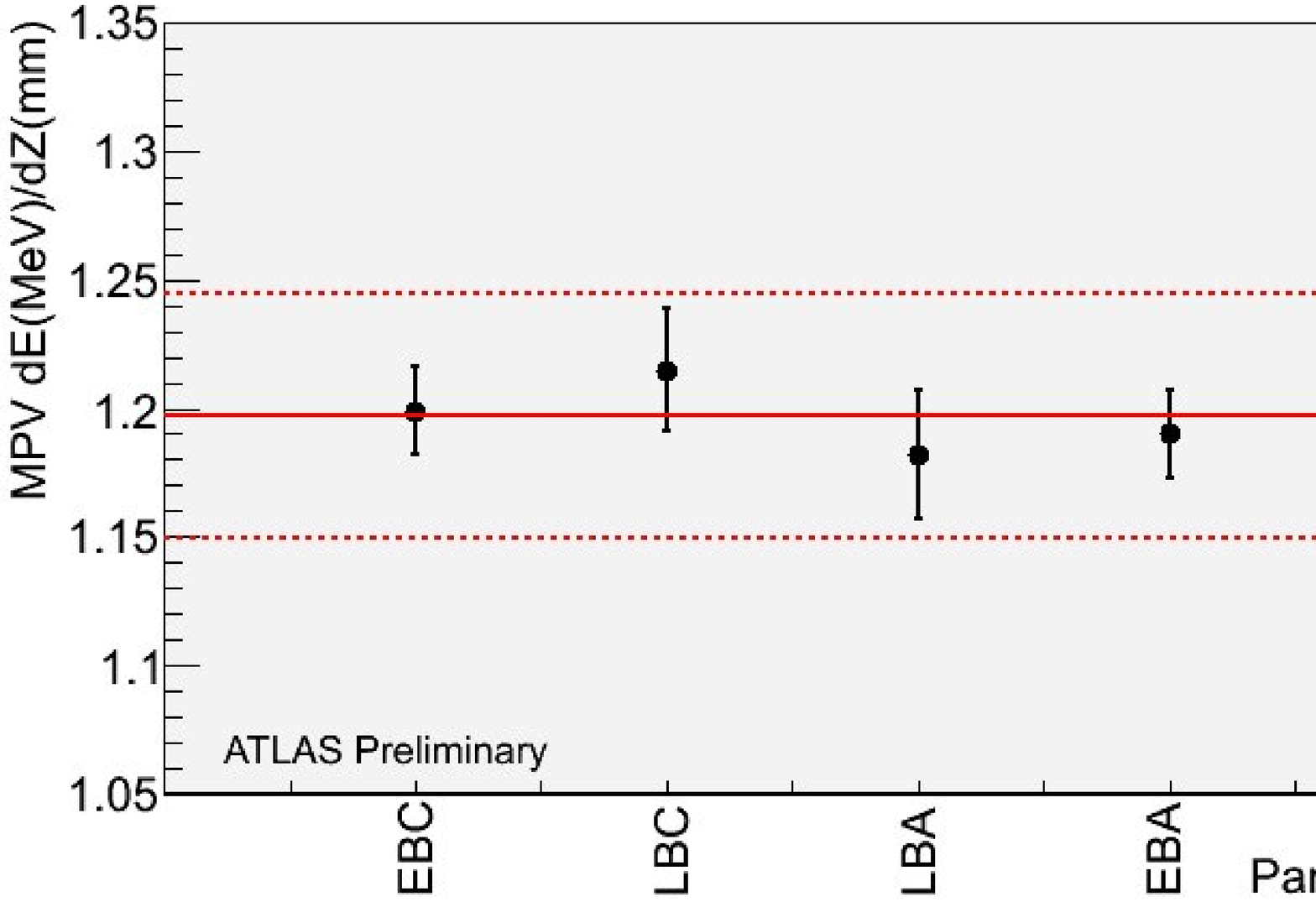}
\caption{Uniformity of the calorimeters: (left) the energy response MPV
as a function of $\eta$ for LAr Electromagnetic
calorimeter in barrel region. Two data types with different cluster 
algorithms, one Monte Carlo simulation, and geometrical cell depth are shown.
These are normalized to the points in $0.3<\eta<0.4$. 
(right)the most probable dE/dx value per partition (in beam direction) 
obtained with single beam. The detector partitions are calibrated with 
embedded $\gamma$-ray sources.} 
\label{fig:CalUniformity}
\end{figure}
The uniformity and stability in operation are the most important 
performance parameters for calorimeters with in-situ cosmic ray measurements 
before the collision data. 
Figure~\ref{fig:CalUniformity} shows the uniformity measurements 
(left) in LAr EM calorimeter. The most probable value (MPV) followed the calorimeter cell 
depth shape. Good agreement with simulation is obtained at the 2\% level
(right) In the tile hadron calorimeter, the uniformity across the partition
is confirmed to be less than 4\%.
Concerning the stability, various values are monitored during the
operations. For instance, the pedestal variation of EM calorimeter was
confirmed to be  
$\pm1$\MeV over 5 months. The noise distribution of the hadron
calorimeter obtained during single beam run is comparable to the 
one obtained during the cosmic ray measurement. These clearly demonstrate the
readiness of the calorimeters for the collision environment.  
\paragraph{Missing \ET performance}
\begin{figure}
\includegraphics[height=.23\textheight]{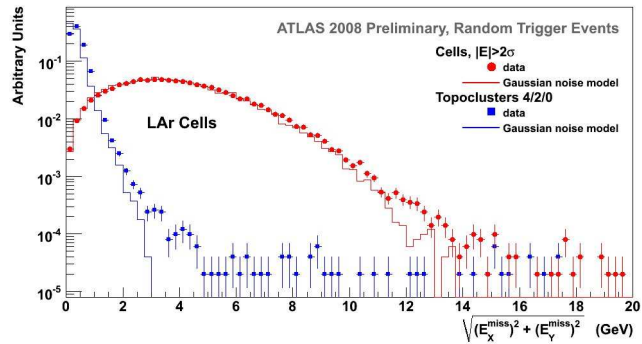}
\caption{\met distributions from random trigger events. Two algorithms 
(red: cell-based, blue: topo-cluster) are shown. Data points represents 
the measurements, while lines illustrate the expected distributions with
Gaussian noise model. Only cells from the LAr EM calorimeter are used in
this plot.}
\label{fig:MetRandom}
\end{figure}
Although the full commissioning of \met performance has to wait 
for the collision data, the tests on noise suppression of the cluster 
algorithms, detector performance checks can be 
carried out at present with a randomly timed trigger. 
Figure~\ref{fig:MetRandom} shows the \met distributions for 
the two main \met calculation algorithms adopted in ATLAS, 
namely (i){\sl the cell based algorithm}; with a simple 
model useful in assessing the basic calorimeter performance, (ii)
{\sl the topological clustering algorithm}; with better
noise suppression and resolution. The tests were carried out with
nearly full detector readout, with 50k events taken with random
trigger. 
The Gaussian noise model (solid histograms) describes the data well, which 
illustrates the degree of understanding of the detector noise at 
the current stage.    
\paragraph{Muon system} 
\begin{figure}
\includegraphics[height=.23\textheight]{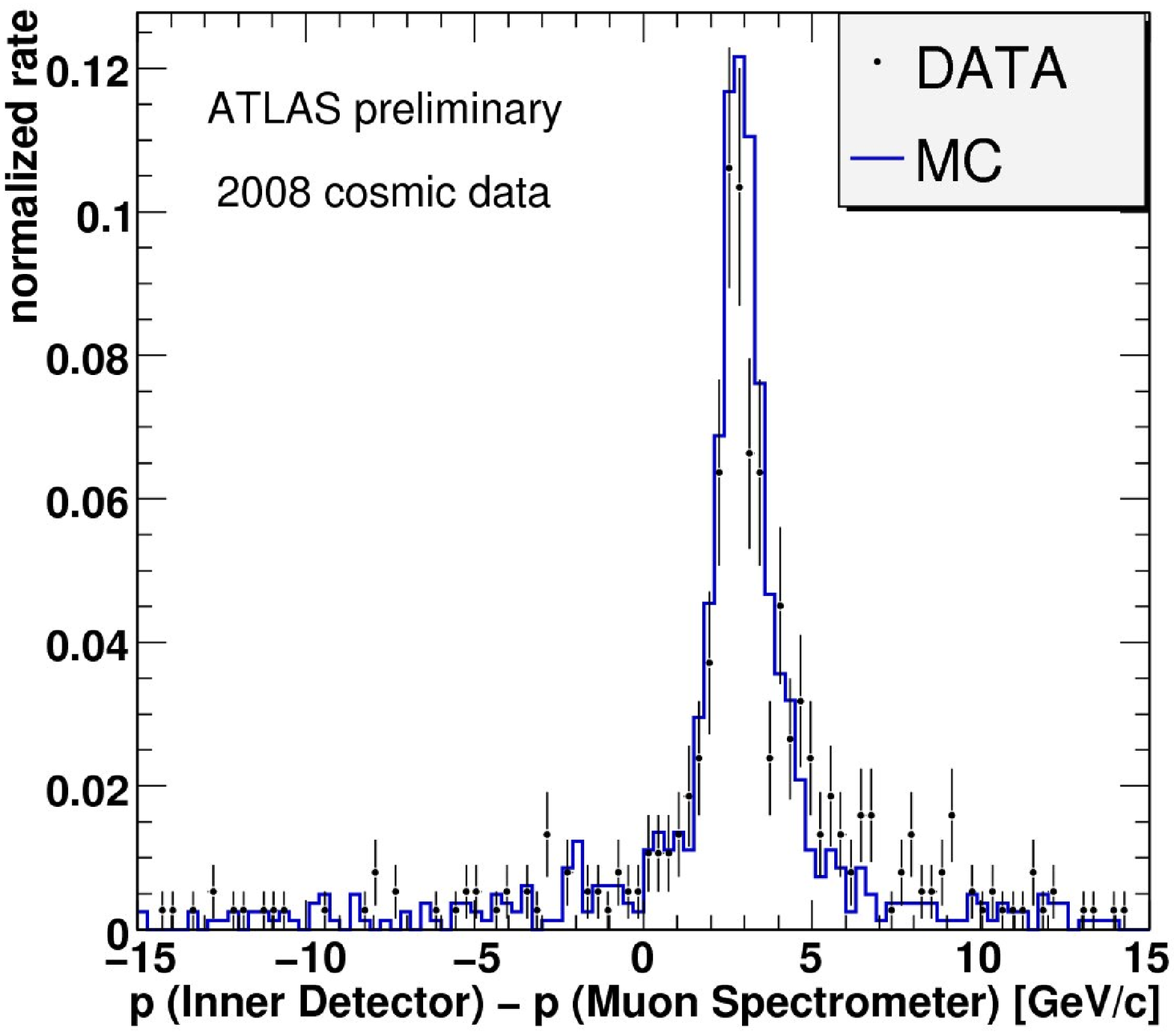}
\includegraphics[height=.23\textheight]{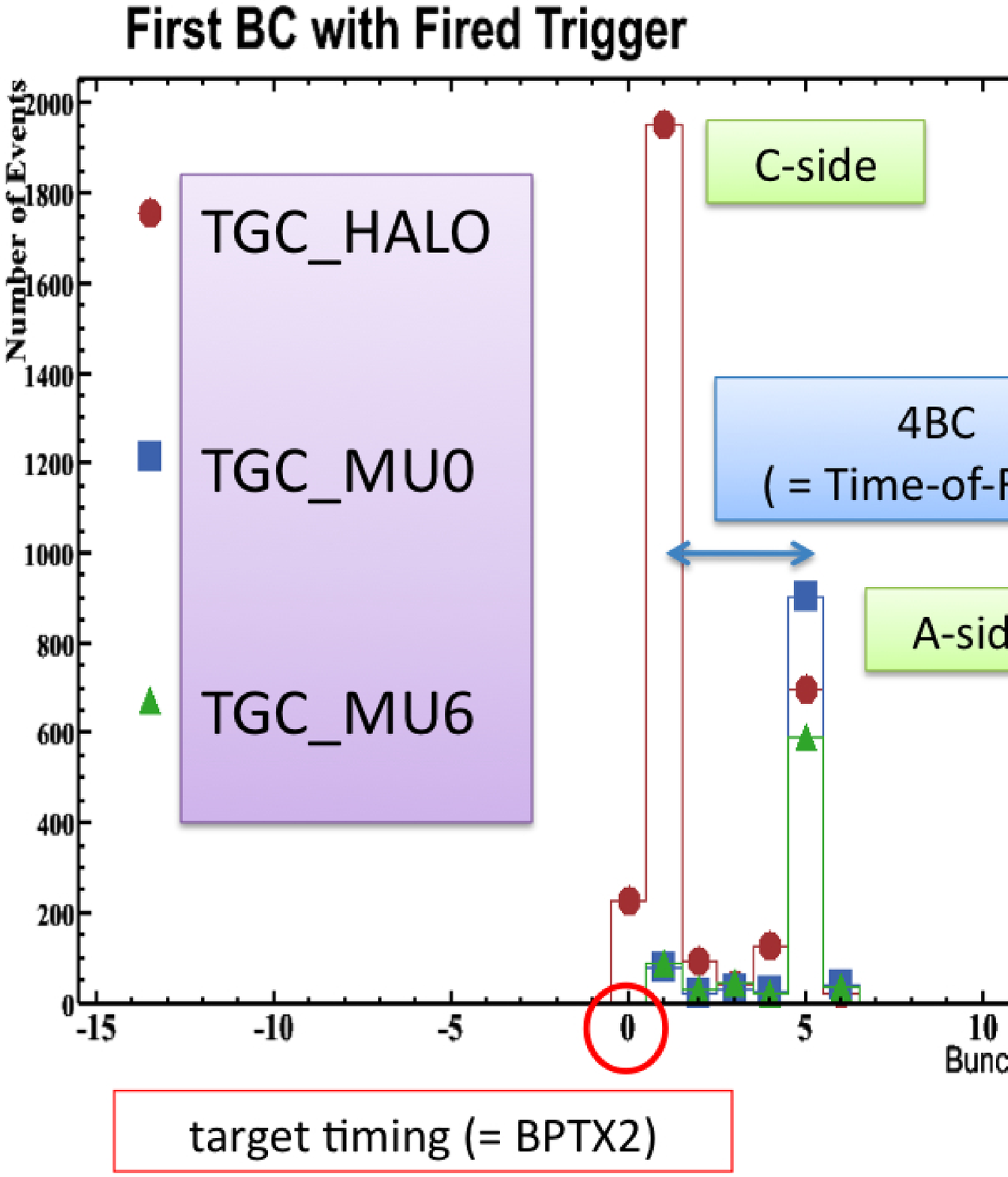}
\caption{(left) difference in cosmic ray momentum measurements between 
inner detector and the muon spectrometer for matched tracks. Expected energy 
loss in Calorimeter (about 3\GeV) and width of data (points) is well described 
by MC simulation (solid histogram). 
(right) LVL-1 muon trigger timing in End-cap region for beam halo events.
The beam halo particles penetrate ATLAS detector from C-side (Geneve side) to
A-side (Mt. Jura side). The event rate of three trigger menus (different 
by matching criteria) are shown. In order to bring A-side to proper 
timing, the timing is shifted by -5 bunch crossing units after this 
measurement.}  
\label{fig:muons}
\end{figure}
The alignment of the muon system in ATLAS is performed with the optical 
system \cite{Muons}. The quality is monitored with the saggita distribution at the 
middle layer of the system. The mean values are fairly close to zero 
($134 (15)\mu m$ for Barrel (End-cap)). Further improvement expected with 
the help of tracking information. 
The cosmic rays penetrating near the interaction point give a good 
opportunity to compare the muon system against the inner tracker. 
Figure~\ref{fig:muons} (left) shows the difference in momentum 
measurements between the two systems. The energy loss and the resolutions 
are well reproduced by the simulation, which is also true for the 
distributions of the differences in the azimuthal angle between the two
systems. Figure~\ref{fig:muons}
(right) is a trigger timing performance plot for the End-cap muon system.
The detector timing (within one bin), relative (between C and A sides) 
and absolute (with respect to the BPTX) timings, are all under good control.

In summary, the systems in ATLAS are all ready for the first 
collision data. More details of the commissioning can be found in another
talk \cite{Commissioning}. 
\section{Strategy of finding SUSY in inclusive search}
The SUSY analysis perspectives presented here are based on the ATLAS
data preparation paper \cite{CSC}. Hence, the analyses assume  
$\sqrt{s}=$14\TeV and $\int{L}=$1\fbinv (Initially the LHC will run at
lower energies, see \cite{PJenni}). 
\begin{figure}
\includegraphics[height=.23\textheight]{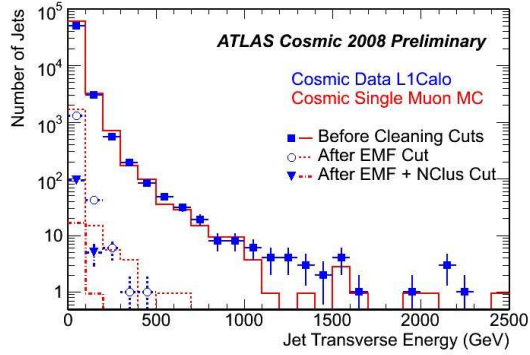}
\caption{The transverse energy distributions of cosmic rays reconstructed 
as Jet. Data and MC simulation are shown with blue square points and red solid
line respectively. Also shown are the data points and the respective 
MC predictions after the event filter for cosmic rays.}
\label{fig:JetDist}
\end{figure}
ATLAS intends to cover all the possible SUSY event topologies, 
that is different NLSP types, number of jets, number of 
leptons, also consideration of the requirements on taus, b-jets, photons, and 
long-lived particles. On top of these, SUSY breaking scale and model
dependent parameters add further complexity. 
Although the phenomenological model independent 
approach is desirable, it is impossible to cover all these without any 
assumption.
We therefore base our analysis strategy on modes where we have a good 
confidence in background estimation. Thus we categorize the event signature 
by number of leptons. Accordingly the major background sources change 
which requires respective estimations.    
\paragraph{0-lepton mode}
Here a control of the huge QCD background is the most important. The
fake \met caused by accelerator, cosmic rays, detector effect/failure
needs to be understood.  
Those local malfunctioning of detector, e.g. noisy/dead cells, or HV
trips, etc, are monitored and rejected event by event basis. 
The bad runs will be dealt with run database. 
Concerning cosmic rays, ATLAS already has a good description of 
data with Monte Carlo simulation as one can confirm in Fig.~\ref{fig:JetDist}.
The cleaning cut on these events, using energy fraction in EM
component, works effectively, which is also reproduced by simulation.
The detector effect on \met caused by the jets falling in poorly 
instrumented region is dealt with the cut on $|\phi^{jet_i}-\phi^{\met}|$,
as the \met is expected to align with the leading \pT jets in such events. 
After all these considerations plus the standard SUSY selections, 
comparable contributions from $t\bar{t}+jets$, $W+jets$, and $Z+jets$ 
processes are expected. However the contribution from
QCD (which current simulation suggest is $<5\%$) has the largest 
uncertainties, where it is difficult to
perform a realistic estimation with Monte Carlo simulation, as
the \met could come from the far tail of the detector response. 
\begin{figure}
\includegraphics[height=.23\textheight]{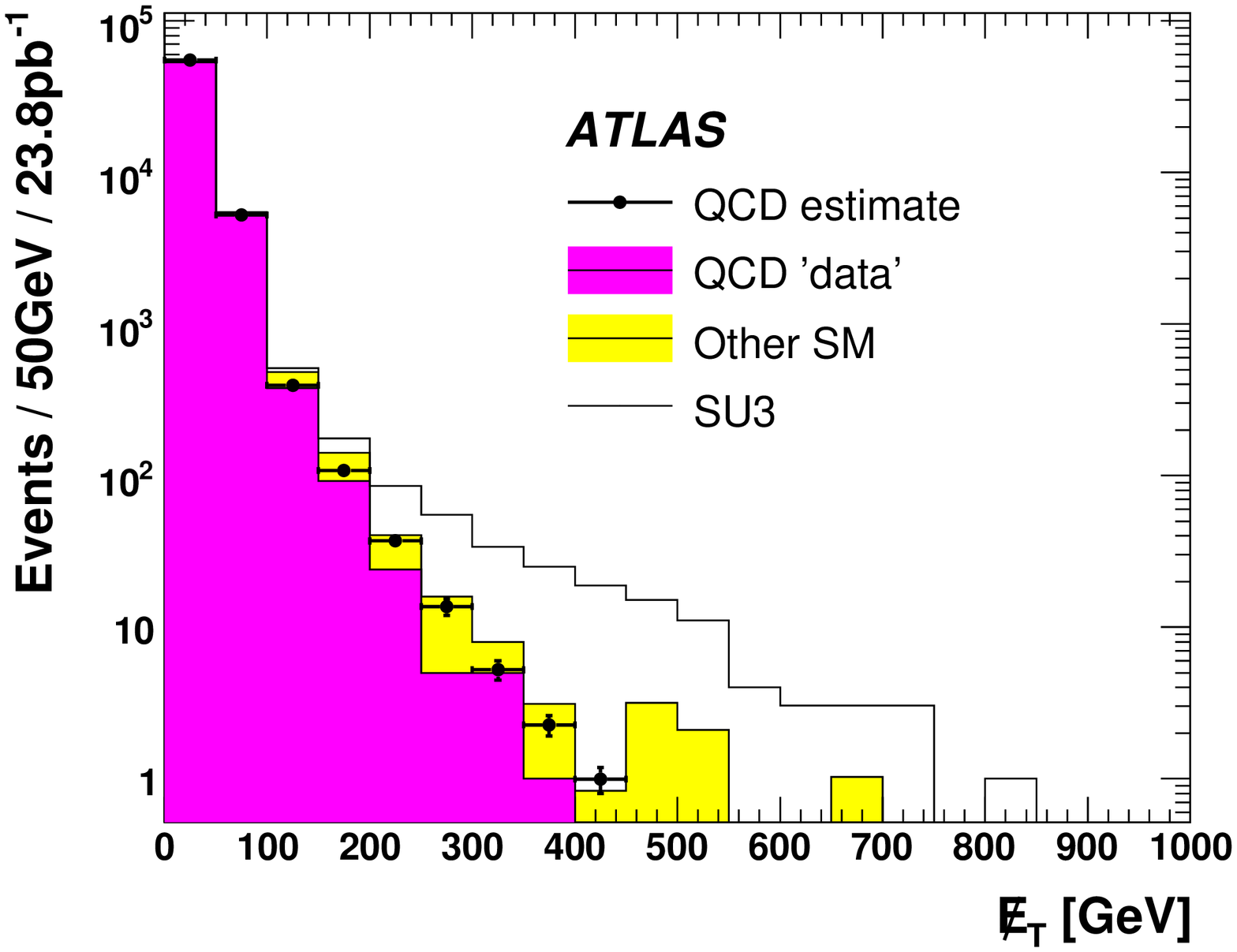}
\includegraphics[height=.23\textheight]{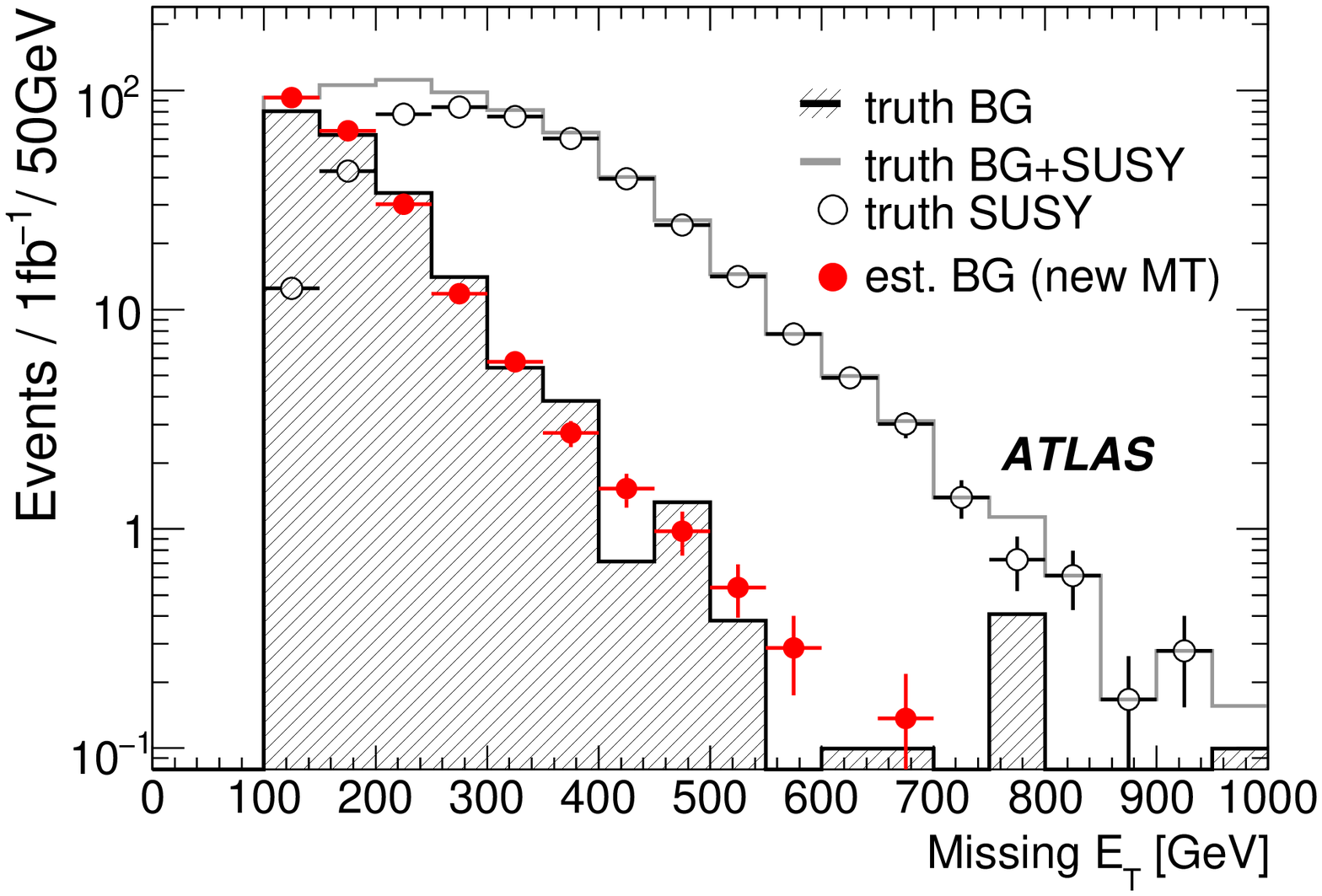}
\caption{Background estimation from the data driven methods: 
(left) \met distributions of SM backgrounds and estimation of QCD contribution 
described in the text after applying 0-lepton mode jet cuts for SUSY analysis. 
Also shown is the distribution of a typical SUSY signal.
(right) \met distributions of SM backgrounds and estimation from the new 
\MT-method. The 1-lepton mode SUSY event selections are applied.}
\label{fig:DataDriven}
\end{figure}
ATLAS developed various procedures to estimates the QCD contributions 
in the SUSY signal regions directly from the collision data (data driven 
estimate). One of which uses three-jet events 
to estimate the \met tail created by jet energy fluctuations \cite{DataDriven}.
Figure~\ref{fig:DataDriven} (left) demonstrates this method. 
Data driven estimates for $Z\rightarrow\nu\nu$, $W$, $t\bar{t}$
have also been developed \cite{CSC}. 
\paragraph{1-lepton mode}
With the requirement of one lepton, in spite of a smaller cross 
section, we can expect better control over many of the backgrounds.
Here we add the isolated lepton requirement and the cut on
transverse mass (\MT). At this point, $t\bar{t}+jets$ is dominant,
while $W+jets$ is important in the high \met tail in the remaining 
background. The QCD background appears to be negligible. 
The data driven technique is developed to 
use the control region (low \MT region) to evaluate the 
normalization and shape of background in signal region, called, 
`\MT method'. 
Further improvement is applied to deal with the case where the SUSY 
signals contaminate the control region (where background would 
be over estimated with naive \MT method). As seen in 
Fig.~\ref{fig:DataDriven} (right) the new method successfully estimates
the size of the background contribution.   
The studies on the 2-lepton mode, higher multi-lepton mode, $\tau$
signature, and with $b$-jets are also carried out in ATLAS
which are reported in the other talks \cite{MultiLepton, TauChannel} .
\paragraph{Inclusive reach in SUGRA parameter scape}  
\begin{figure}
\includegraphics[height=.23\textheight]{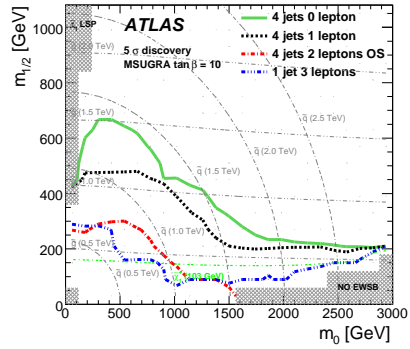}
\caption{Discovery reach of inclusive SUSY search in mSUGRA parameter 
space assuming $\sqrt{s}=$14\TeV and 1\fbinv integrated luminosity.}
\label{fig:DiscoveryReach}
\end{figure}
After all these considerations, the SUSY discovery reach for 14\TeV, 1\fbinv
is obtained, taking into account the expected uncertainties on SM backgrounds 
using 1\fbinv of integrated luminosity \cite{DiscoveryReach}. The systematic errors are estimated to be $50\%$
on QCD processes, and $20\%$ on $t\bar{t}$, $W$, $Z$+jets. This is 
illustrated in Fig.~\ref{fig:DiscoveryReach}, where 4 jets plus different
lepton requirements are shown. It is also seen that multiple signatures 
are expected in most of the parameter space, meaning the redundant 
analysis is possible. 

With the 10\TeV and 100\pbinv, lower cross sections and 
worse background control are anticipated, which would naturally degrade 
the reach. However it is still expected to go beyond Tevatron limit.  
Exclusive measurements are not covered in this review. The reader is 
referred to other dedicated talk \cite{Exclusive}. 
\section{SUSY with special signatures}
\begin{figure}
\includegraphics[height=.23\textheight]{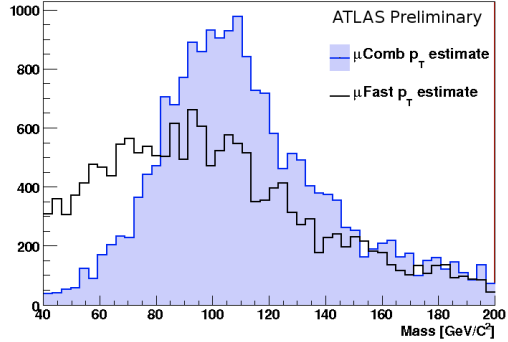}
\caption{The mass distribution obtained at the LVL-2 trigger stage.
$M_{\tilde{\ell}}=$100\GeV, 500\pbinv luminosity is assumed. Cuts 
of $\pT>$40\GeV, $\beta_{measured}<0.97$ are applied. Background 
from the inclusive single muon events -not shown here- has a tail 
up to 80\GeV in mass, 
which can be effectively eliminated in this case.}
\label{fig:LongLived}
\end{figure}
In addition to the SUSY signatures with jets and \met, 
ATLAS will search for the various well-motivated physics, 
which predict characteristic signatures e.g. GMSB, AMSB, 
Split SUSY, RPV, etc. Regardless of the models considered,
the signatures can be generally categorized in terms of NLSP type and 
lifetime. In each category, the relevant detector component
is different. In most of the models, the signal has basis 
SUSY characteristics such as multi high-\pT jets, large \met, thus 
such signals has high efficiency with standard trigger menu for SUSY. 
The trigger menu based on these special signatures adds  
redundant triggers. In some cases, the dedicated menu and algorithm
have to be prepared for cases, when the basis SUSY 
signature is missing.

One such scenario is the additional high-\pT photons in GMSB 
models with a $\tilde{\chi}^0_1$ NLSP. There we expect high-\pT
jets, large \met, then 2 high-\pT photons, hence we expect 
very small SM backgrounds. With a good control over 
the fake photons, the potential for early discovery is high. 
In GMSB models, NLSP could be long-lived, then the photon no
longer points back to the interaction point. ATLAS is equipped with 
finely granulated EM calorimeter layers in $\eta$ direction, and also 
these have a good timing resolutions. Hence the 
detector is sensitive to `non-pointing' photons and the lifetime of the NLSP 
could be measured \cite{non-pointing}. 

Another scenario is heavy meta-stable charged particles 
which penetrate through ATLAS rather like high-\pT muons \cite{sleptons}. 
Uncolored particles, such as sleptons, can be distinguished from muons
since they would have small $\beta$ and have
significant delay in TOF. Special trigger and reconstruction
algorithm are essential to detect such particles.  In ATLAS,
special menu is prepared to select the low $\beta$, high mass
particles at the LVL-2 trigger stage, as seen in Fig.~\ref{fig:LongLived}.

At Event Filter (LVL-3 trigger), finer mass reconstruction 
is possible, where the mass resolution for 100\GeV slepton would be 
16\%. 
Yet another unique signature is foreseen with so called `R-hadrons'
with color interactions in detector. For these one would expect
charge flips inside calorimeter, invisible tracks inside 
inner trackers, etc \cite{R-hadrons}.   
There are other unique signatures which ATLAS will pursue.
(One such scenario is R-parity violation \cite{Rparity})  
\section{Conclusions and outlook}
ATLAS is finishing its final tests on detectors, DAQ, and online 
systems using cosmic rays and single beam data taken during 2008/2009. 
Analysis techniques in estimating the backgrounds 
using real data have also been developed and are maturing for practical 
use, still new ideas and improvements are evolving.
We confirmed that there is a good potential of discovery at 10\TeV,
100-200\pbinv. Also preparations for SUSY signals accompanying 
special signatures are ready. 

We look forward to bringing the big news and surprises in next year.




\begin{theacknowledgments}
The author wishes to thank all the members of ATLAS collaboration for 
the excellent results used in this talk. 
\end{theacknowledgments}



\bibliographystyle{aipproc}   




\end{document}